\documentclass[conference]{IEEEtran}
\IEEEoverridecommandlockouts
\usepackage{cite}
\usepackage{amsmath,amssymb,amsfonts}
\usepackage{algorithmic}
\usepackage{graphicx}
\usepackage{textcomp}
\usepackage{xcolor}
\def\BibTeX{{\rm B\kern-.05em{\sc i\kern-.025em b}\kern-.08em
    T\kern-.1667em\lower.7ex\hbox{E}\kern-.125emX}}
\begin{document}

\title{[DC] bRight XR: How to train designers to keep on the bright side?}

\author{\IEEEauthorblockN{Romain Rouyer}
\IEEEauthorblockA{\textit{Université de Lorraine} \\
\textit{UR7312 PErSEUs}\\
Metz, France \\
romain.rouyer@univ-lorraine.fr}
\and
\IEEEauthorblockN{David Bourguignon}
\IEEEauthorblockA{\textit{Université de Lorraine} \\
\textit{UR7312 PErSEUs}\\
Metz, France \\
david.bourguignon@univ-lorraine.fr}
\and
\IEEEauthorblockN{Stéphanie Fleck}
\IEEEauthorblockA{\textit{Université de Lorraine} \\
\textit{UR7312 PErSEUs}\\
Metz, France \\
stephanie.fleck@univ-lorraine.fr}
}

\maketitle

\begin{abstract}
This research project aims to promote ethical principles among designers engaged in adaptive-XR by providing tools for self-assessment. We introduce a Design-Based Research (DBR) methodology to build bRight-XR, a framework including a heuristic evaluation matrix and based on learning theory.
\end{abstract}


\section{Introduction}

Researchers in human-computer interactions (HCI) are concerned with ethical tensions in persuasive design practices. The boundaries between manipulative mechanisms and positive technologies are generally thin (e.g.: deceptive pattern \cite{brignullDeceptivePatternsUser2023}). 
The use of standard digital interfaces highlights designers' critical role and responsibility in ensuring users' integrity and well-being.

In a context of democratization of immersive and body-adaptive technologies (adaptive-XR), warning signals from philosophers \cite{madaryRealVirtualityCode2016}, designers \cite{tsengDarkSidePerceptual2022} and educators \cite{leeE3XRAnalyticalFramework2021} insist on the need for new approaches to design practices. The associated findings aim to limit the negative consequences of embodied design \cite{abrahamsonEmbodiedDesign2019} and urge designers to respect their users' short-  and long-term well-being. They encourage a paradigm shift to a fair and virtuous user experience that benefits the user and that goes beyond the search for experiences that are merely pragmatic or hedonistic. 

Knowing this and recommending this, however, does not empower designers to make these changes. Given these concerns, we aim to develop and evaluate bRight-XR, a heuristic framework that incorporates ethical dimensions, which can be used both as a design guidance tool and as a training tool for adaptive-XR designers.

In this paper, we introduce the proposed design-based research (DBR) methodology for designing such a training toolkit, mobilizing in a three-part plan: (1) Design of a heuristic recommendation grid, (2) Test of pedagogical prototypes on the ground and (3) Pre-validation of a training kit.

\section{Design-based research method for bRight-XR}

Our research aims to bridge the gap between research and practice in adaptative-XR design. Design-based research (DBR) was heralded as a practical research methodology that could effectively bridge this gap with the support of learning theory \cite{andersonDesignBasedResearchDecade2012}. This paper describes how we propose to mobilize an ergonomic process and co-design activities on the ground with researchers and designers.

\subsection{Design a heuristic evaluation matrix}

The process of creating a heuristic evaluation grid involves identifying distinct dimensions, tying them to usability standards and a scale for evaluation. There are several possible approaches (e.g.: empirical, deductive, etc.), each with different effects and scopes, as Pineau and Fabre emphasize in the construction of their captology analysis matrix \cite{pineauEvaluerCaptologieDesign2021}.

In our context, we will first conduct a systematic review of the scientific literature to highlight existing theoretical frameworks \cite{leeE3XRAnalyticalFramework2021}, guidelines \cite{madaryRealVirtualityCode2016, fuchsCharteRecommandationsUsage2019, meachamGuidelinesEthicsDesign2022} and design process \cite{wangUnderstandingVirtualDesign2024}.
To complement with an empirical approach, a series of semi-directive interviews with designers, researchers and educators expert in XR will allow us to identify feedback and projections from the ground. 

Then, our data analysis will enable us to identify and qualify the main dimensions currently implemented in a matrix containing, but not limited to: disciplinary fields, type of interaction (input, output), modalities of measurement (e.g.: posture, gaze, voice, facial/labial expression, gesture, electroencephalography signal, heart rate), temporality of the effect (before, during, temporary, persistent), the reliability indicator (number of studies/observations reported), the robustness of the usage (massive, emerging, niche).

This matrix will enable us to identify and propose a list of heuristic criteria, associated with an evaluation scale, considering the ethics of care. These criteria will be prioritized using a hybrid evaluation: a workshop for the qualitative dimension and a survey to statistically validate the results.

\subsection{Test pedagogical prototypes}

To evaluate the analysis matrix, we will apply the scoring method to different pedagogical scenarios, using experimental prototypes in XR-adaptive. This phase will enable us to define training leverages for designers.

The scenarios and prototypes will be produced in collaboration with pedagogical engineers, designers and technologists, using the Design Fiction approach. We will partly refer to prospective scenarios that have been developed in recently published studies \cite{eghtebasCoSpeculatingDarkScenarios2023, greenUsingDesignFiction2019, wangDarkSideAugmented2023}.

\subsection{Pre-validation of a training kit}

The outcomes of the Design Fiction investigation will be incorporated into a demonstration and educational toolkit. The kit will consist of a heuristic criteria evaluation grid and a set of recommendations. A concluding preliminary investigation will be conducted to evaluate its impact on designers, which will be validated through the utilization of a questionnaire.

The implementation of this questionnaire will be influenced by the previous work of Bastien and Scapin, \cite{bastienEvaluatingUserInterface1994}, which focused on usability, and the work of Fleck et al., which focused on emerging educational technologies \cite{baraudonConceptionEchelleFrancaise2021}, and the consideration of long-term well-being \cite{watermanQuestionnaireEudaimonicWellBeing2010}.

This kit will be fully documented, published as an Open-Source project and built on open science standards to enable interoperability with other initiatives and welcome contributions.

\section{Discussion}

This project addresses the emerging nature of XR-adaptive technology and associated design techniques, which are evolving rapidly. It presents a unique opportunity to propose a comprehensive and ethical design framework at an early stage, yet it presents challenges in evaluating it in the absence of widespread implementation.

Furthermore, generative techniques (e.g., AI and Large Language Models) are soon to enter design arsenals, making it even more challenging for a designer to comprehend ethical aspects.

The uniqueness of this research project is derived from its multidisciplinary nature and its intention to establish a dialogue among diverse research fields that have, each in their own way, been accumulating cutting-edge knowledge and expertise for decades. These include cognitive sciences (psychology, philosophy, sociology), science and technology studies (signal processing, sensors, on-board electronics), and educational sciences.

\section{Conclusion}
To conclude, we anticipate demonstrating that a planned and heightened emphasis on ethics training on adaptative-XR designer, at an early stage of the democratization of this technology, could mitigate the adverse effects that digital technologies can have on individuals. 

To fully exploit the performance of such a pedagogical toolkit, suitable dimensions need to be identified. The robustness of the approach appears to be attributed to its capacity to span multiple disciplinary boundaries and take into account a variety of user-risk scenarios.


\end{document}